\documentclass[%
 aip,
 amsmath,amssymb,
 reprint,%
]{revtex4-1}
\usepackage{multirow}
\usepackage{graphicx}
\usepackage{dcolumn}
\usepackage{bm}

\usepackage[utf8]{inputenc}
\usepackage[T1]{fontenc}
\usepackage{mathptmx}
\usepackage{etoolbox}

\usepackage{subfig, graphicx}
\usepackage{caption}
\makeatletter
\def\@email#1#2{%
 \endgroup
 \patchcmd{\titleblock@produce}
  {\frontmatter@RRAPformat}
  {\frontmatter@RRAPformat{\produce@RRAP{*#1\href{mailto:#2}{#2}}}\frontmatter@RRAPformat}
  {}{}
}%
\makeatother
\begin{document}

\preprint{AIP/123-QED}

\title[]{Efficient aerodynamic coefficients prediction with a long sequence neural network}
\author{A. Zemin Cai}
 \affiliation{Key Laboratory of Digital Signal and Image Processing of Guangdong Province, Shantou, China}
 \affiliation{ 
 	The Department of Electronic Engineering, Shantou University, Shantou, China
 }%

\author{B. Zhengyuan Fan}%
\affiliation{ 
The Department of Electronic Engineering, Shantou University, Shantou, China
}%

\author{C. Tianshu Liu}

\altaffiliation[Authors to whom correspondence should be addressed: ]{tianshu.liu@wmich.edu}

\affiliation{%
Department of Mechanical and Aerospace Engineering
Western Michigan University, Kalamazoo, MI, USA
}%

\date{\today}

\begin{abstract}
Traditionally, deriving aerodynamic parameters for an airfoil via  Computational Fluid Dynamics requires significant time and effort. However, recent approaches employ neural networks to replace this process, it still grapples with challenges like lack of end-to-end training and interpretability. A novel and more efficient neural network is proposed in this paper, called AirfoilNet. AirfoilNet seamlessly merges mathematical computations with neural networks, thereby augmenting interpretability. It encodes grey-scale airfoil images into a lower-dimensional space for computation with Reynolds number, angle of attack, and geometric coordinates of airfoils. The calculated features are then fed into prediction heads for aerodynamic coefficient predictions, and the entire process is end-to-end. Furthermore, two different prediction heads, Gated Recurrent Unit Net(GRUNet) and Residual Multi-Layer Perceptron(ResMLP), designed to support our iteratively refined prediction scheme. Comprehensive analysis of experimental results underscores AirfoilNet's robust prediction accuracy, generalization capability, and swift inference.
\end{abstract}

\maketitle

%

\section{\label{sec:introdection}INTRODUCTION}

The design of spacecraft necessitates a comprehensive knowledge of their intricate aerodynamics. Yet, acquiring data on the flow field and aerodynamic coefficients around an airfoil remains a challenging endeavor. Conventional methods, including wind tunnel experiments and Computational Fluid Dynamics (CFD) simulations, demand extensive lead times and considerable labor, adding to the intricacy of the process. Wind tunnel experiments and CFD simulations are tightly linked to the Navier-Stokes equations. Presently, this purely scientific approach demands significant computational resources when facing a wider array of airfoils and more complex conditions. While certain methodologies\cite{pod,dmd} have endeavored to enhance computational efficiency, they have yet to offer viable solutions for multifaceted, transient, and disparate scenarios. With the advancements in machine learning, numerous machine learning techniques\cite{ai1, ai2, ai3} have found applications within the field of aerodynamics, and this new paradigm has greatly contributed to aerodynamic research. Nevertheless, certain approaches continue to grapple with excessive human intervention, exemplified by Zuo et al.'s work\cite{fast}.

The neural network Zuo et al.\cite{fast} proposed is to predict flow fields around the airfoil. This network comprises two components: a feature extractor, leveraging Vision Transformer(VIT)\cite{vit}, dedicated to extracting airfoil image features, and another segment tasked with estimating the flow field. The crux of the problem is in the first part where they need to pre-train a VIT to encode airfoil images as prior knowledge.
 On the contrary, this step is omitted in our approach, achieving genuine end-to-end training and prediction.
 
Nevertheless, the majority of existing methods for estimating airfoil aerodynamic parameters rely heavily on convolutional neural networks(CNNs)\cite{cnn} due to their higher efficiency. Zhang et al.\cite{application} considered employing a multi-layer CNN to predict lift coefficients for various airfoil shapes, given parameters such as airfoil shape, angle of attack(AOA), Mach, and Reynolds number. There are many similar methods: Sun et al.\cite{sun2021} constructed a deep neural network designed to intake geometric parameters and airfoil coordinates as inputs and outputs pressures and velocities of corresponding coordinates. Sekar et al.\cite{fast2019} extracted features from the airfoil images by using CNNs, which were then fed into a multilayer perceptron(MLP) along with the Reynolds number and angle of attack for flow field prediction. Our approach draws inspiration from Sekar et al.\cite{fast2019}, albeit with a distinct approach: rather than directly utilizing the concatenation of CNN’s output and the aerodynamic parameters as the subsequent input, we independently encode them through embedding layers. This process generates multiple different data distributions, which are accumulated to form a new distribution, serving as the input for the subsequent step. Obviously, deep learning models find it easier to learn simpler data distributions compared to complex ones. Hence, our method is notably more efficient.

\begin{figure*}[htbp]
	\centering
	
	\includegraphics[width=0.7\linewidth]{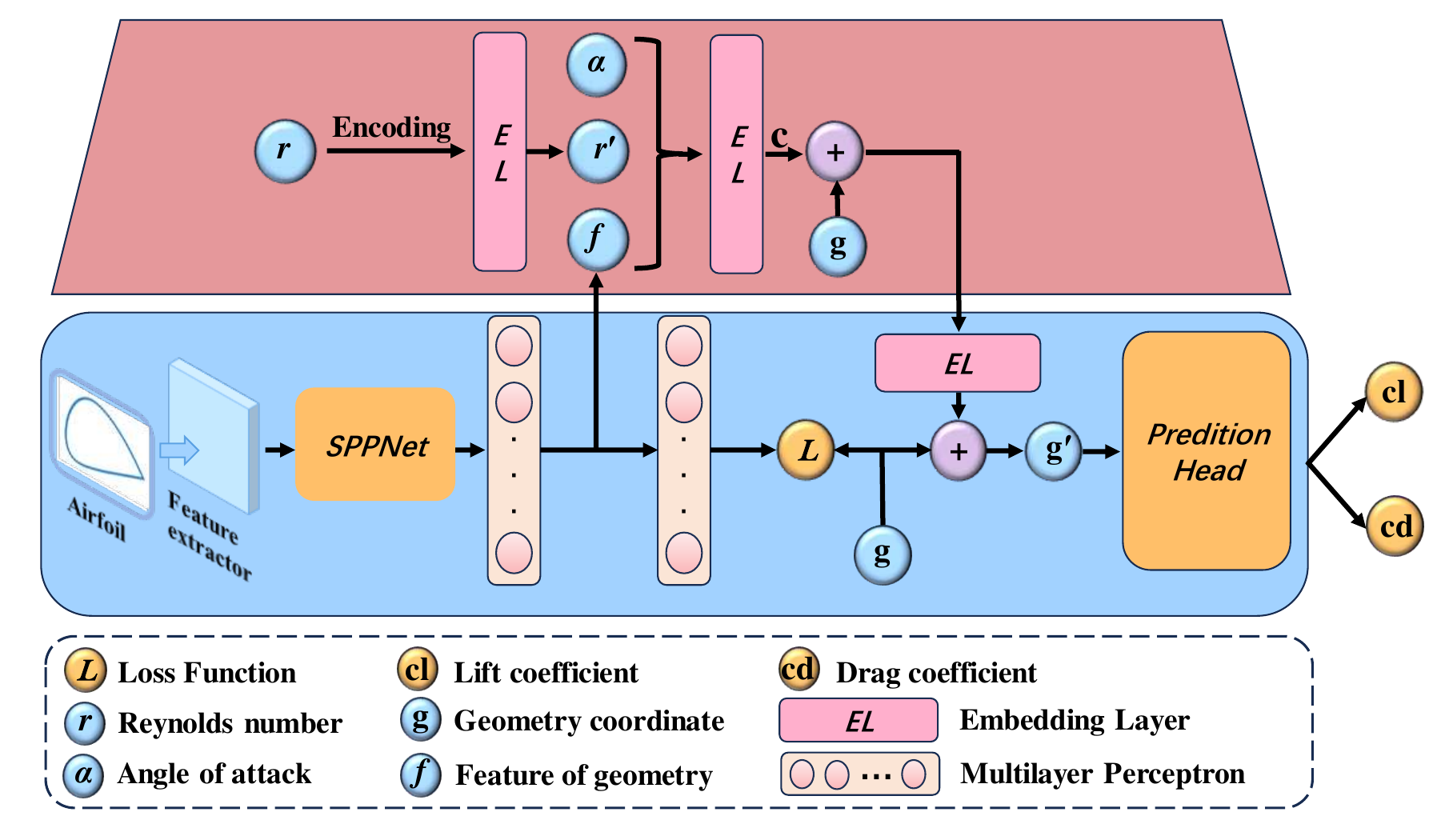}
	
	\caption{Overview of AirfoilNet.}
	
	\label{fig:airfoilnet}
\end{figure*}

In our research, we try to break through the shortcomings of current deep learning methods in physical field. The first is that most deep learning approaches predict aerodynamic parameters rely on MLP. However, this frequently necessitates expanding the MLP to increase depth and width for precise accuracy. For example, the prediction head has 1042K tranable parameters in Zuo et al.'s method\cite{fast}, while ours has only 200K, where ResMLP has 120K and GRUNet\cite{gru} has 80K. The second is removing the process of manual interference. This issue primarily arises from the independent feature extraction operation. The independence is a result of larger parameters or intricate structures of feature extractor, consequently elongating the model's inference time. In contrast, our approach has a simple CNN-based feature extractor consisting of only five convolutional layers with only 4270 parameters. We believe that for airfoil images, there exists more than 90\% redundant information, so the feature extractor is not as crucial. Finally, we want to make our deep learning model more interpretable, where every operation in our method is interpretable. In summary, our contributions are as follows:
\begin{itemize}
	\item An end-to-end and efficient deep learning model is proposed for aerodynamic coefficient prediction called AirfoilNet. This model features two prediction heads: the Residual Multi-Layer Perceptron(ResMLP) and Gated Recurrent Unit Net(GRUNet). They are versatile, allowing for independent or combined prediction. 
	
	\item A new prediction strategy called iterative refinement is proposed in this paper. This is designed for GRUNet. During each iteration of the GRUNet's learning process, it focuses on the change in $\Delta C$, progressively converging the results towards the true value through iterative accumulation.
	
	\item We introduce a novel Reynolds number normalization method that addresses the challenge of sparse results caused by traditional methods. Particularly, when the difference between the maximum and minimum Reynolds number is substantial.
	
	\item We provide the code, as well as the trained weights file on Github repository\cite{github}. Allow researchers to reproduce, improve, and extend our work.
\end{itemize}

\section{METHODOLOGY}

Fig.~\ref{fig:airfoilnet} provides the overall architecture of AirfoilNet. The system initially acquires airfoil image features using a feature extractor, these features then undergo spatial pyramid pooling network(SPPNet)\cite{sppnet} processing. This ensures uniformity in airfoil images of varied sizes, providing remarkable scalability to the neural network. The obtained feature map is mapped into a feature vector by a multilayer perceptron, which is operated with other aerodynamic parameters and finally fed into the prediction head to find the lift and drag coefficients. This section presents the details of our proposed method.

\subsection{Reynolds number Encoding}
\label{sec::encoding}
The Reynolds number stands out among other aerodynamic parameters due to its significantly larger magnitude. In deep learning, we must normalize it to ensure the stability of gradient computations. The current methods all use the following approach to normalize the Reynolds numbers:
\begin{equation}
	\frac{r_i-r_{\text {avg }}}{\max (r)-\min (r)}
\end{equation}
where $r$ is all the Reynolds numbers and $r_i$ is one of them.

\begin{figure*}[htbp]
	\centering
	\begin{minipage}[b]{1\textwidth}
		\subfloat[]{\includegraphics[width=0.33\textwidth]{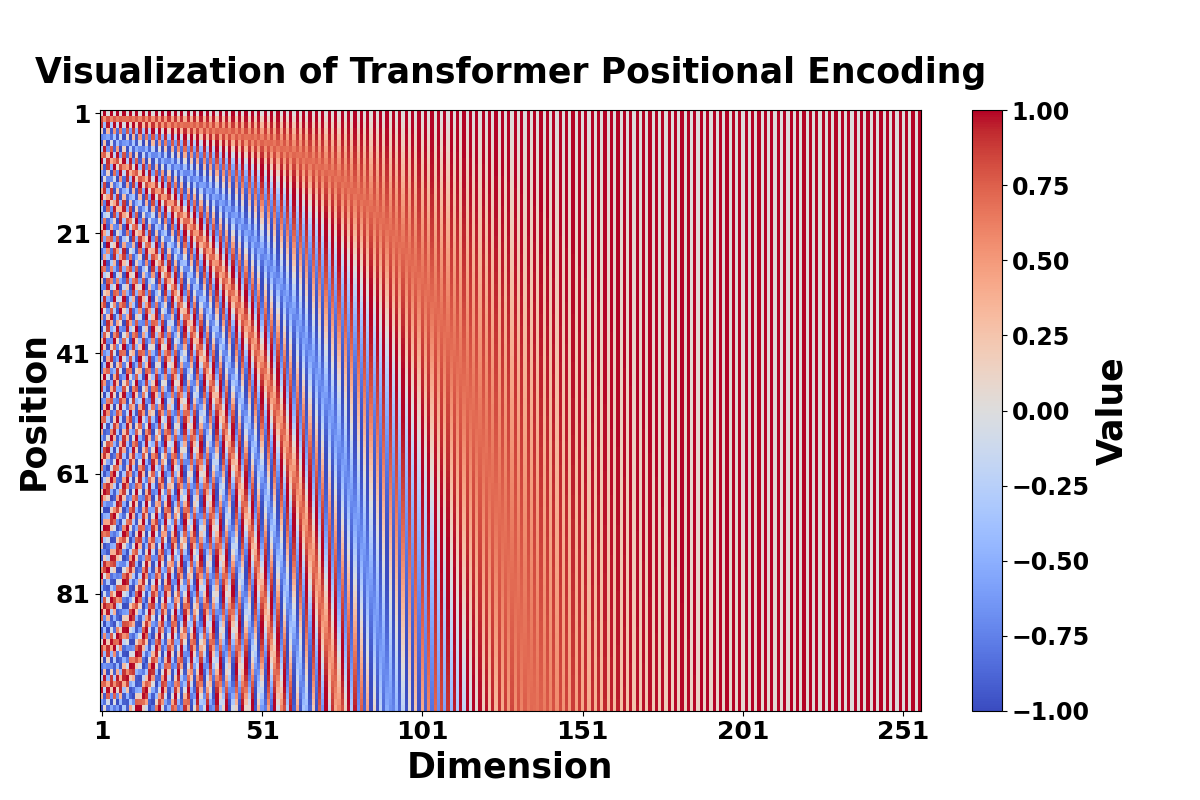} }
		\subfloat[]{\includegraphics[width=0.33\textwidth]{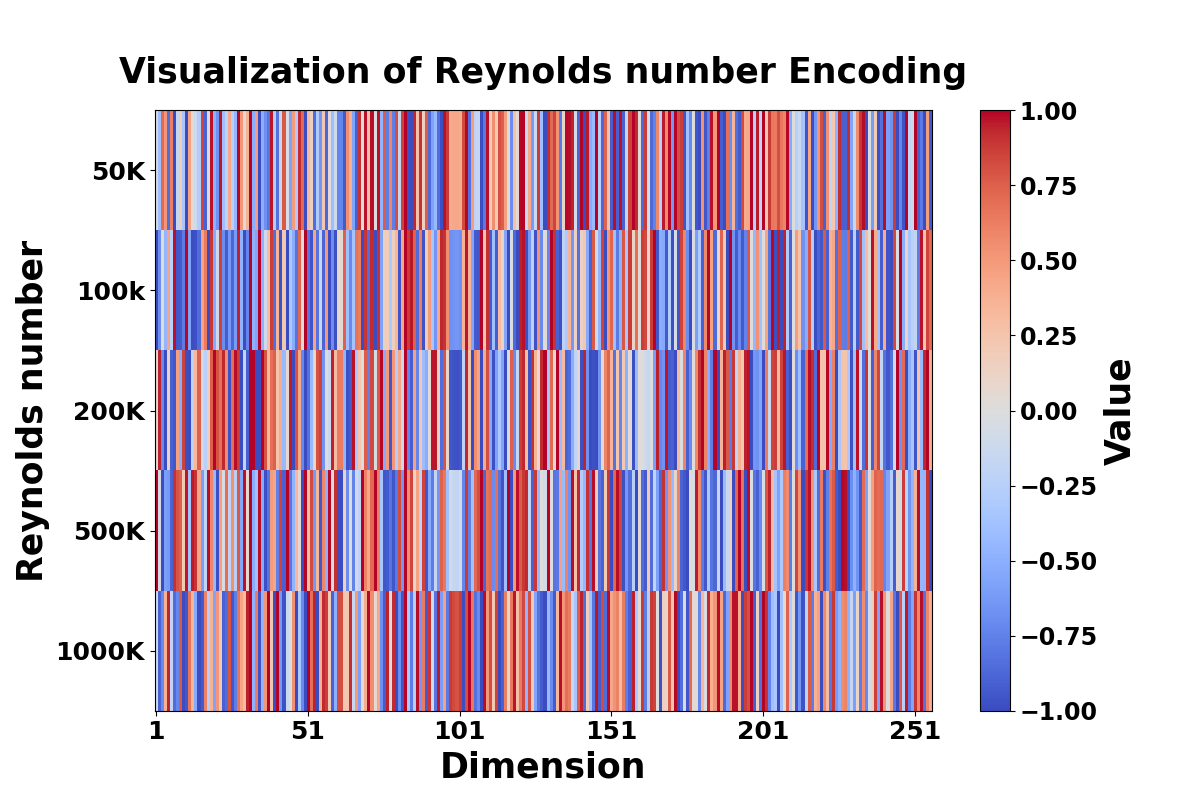}} 
		\subfloat[\label{c}]{\includegraphics[width=0.33\textwidth]{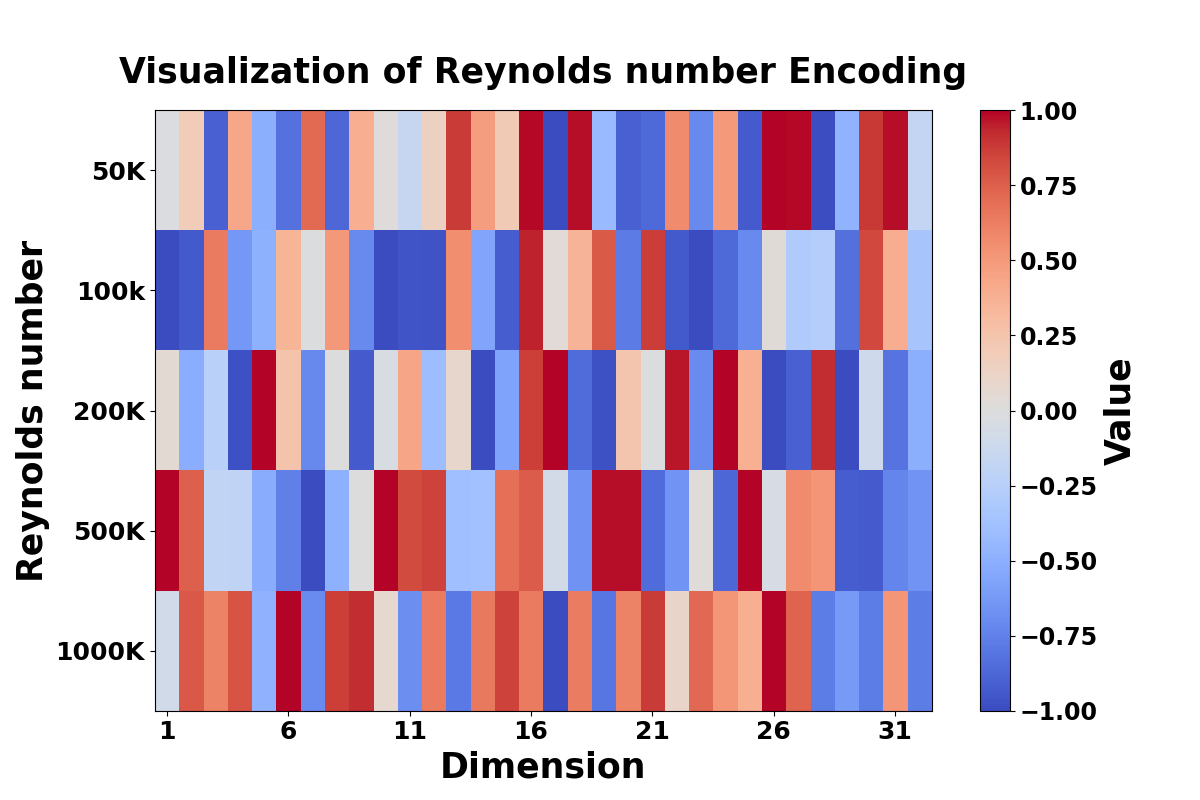}} 
	\end{minipage}
	
	\captionsetup{singlelinecheck=false, justification=raggedright}
	\caption{Visualization of positional encoding and Reynolds number encoding. (a) denotes the encoding of word positions in word vectors in Transformer, (b) and (c) indicate that different Reynolds numbers are encoded into different vectors and the values in the vectors are from -1 to 1.}
	
	\label{fig:encoding}
\end{figure*}

If Reynolds number is normalized using this method, the final outcome tends to be sparse, particularly when there's a substantial gap between the minimum and maximum Reynolds numbers. Therefore, we propose a novel approach to map the Reynolds number into higher dimensions in frequency bands, enhancing the distinct feature representation among different Reynolds numbers:
\begin{equation}
	\left\{
	\begin{aligned}
		& \alpha=r \cdot \exp \left({2 i}/{d} \cdot \log (n)\right) \\
		& \bar{r}=\operatorname{cat}([\cos \alpha, \sin \alpha])
	\end{aligned}
	\right.
\end{equation}
where $r$ is Reynolds number, $d$ is the dimension of the encoded Reynolds number, which is set to 32 in our experiments. $i$ is dimension index, from 0 to $d-1$, $n$ represents the number of Reynolds number categories, and $\bar{r}$ denotes the encoded Reynolds number. cat$(\cdot)$ is a function that combines the inputs. This encoding is inspired by the positional encoding in the Transformer\cite{Transformer}. Fig.~\ref{fig:encoding} provides a visualization of these encoding methods.

Outlined in our approach is the mapping of the Reynolds number from sinusoidal and cosine functions to higher-dimensional vectors. This transformation ensures the uniqueness of each Reynolds number, eliminating any sparsity that might cause the model gradient to vanish.

\subsection{Spatial Pyramid Pooling Network}
Given an RGB image with shape $H \times W \times 3$, where $H$ and $W$ are the height and width of the image, the '3' refers to the channels of the image corresponding to R, G and B. In our method, the image is extracted as a feature map of shape $H/32 \times W/32 \times C$. Notably, varied sizes of input images produce distinct outcomes. However, the neuron count is fixed in the linear layer, so standardizing the feature map shape to a consistent value becomes imperative. 

\begin{figure}[htbp]
	\centering
	
	\includegraphics[width=0.9\linewidth]{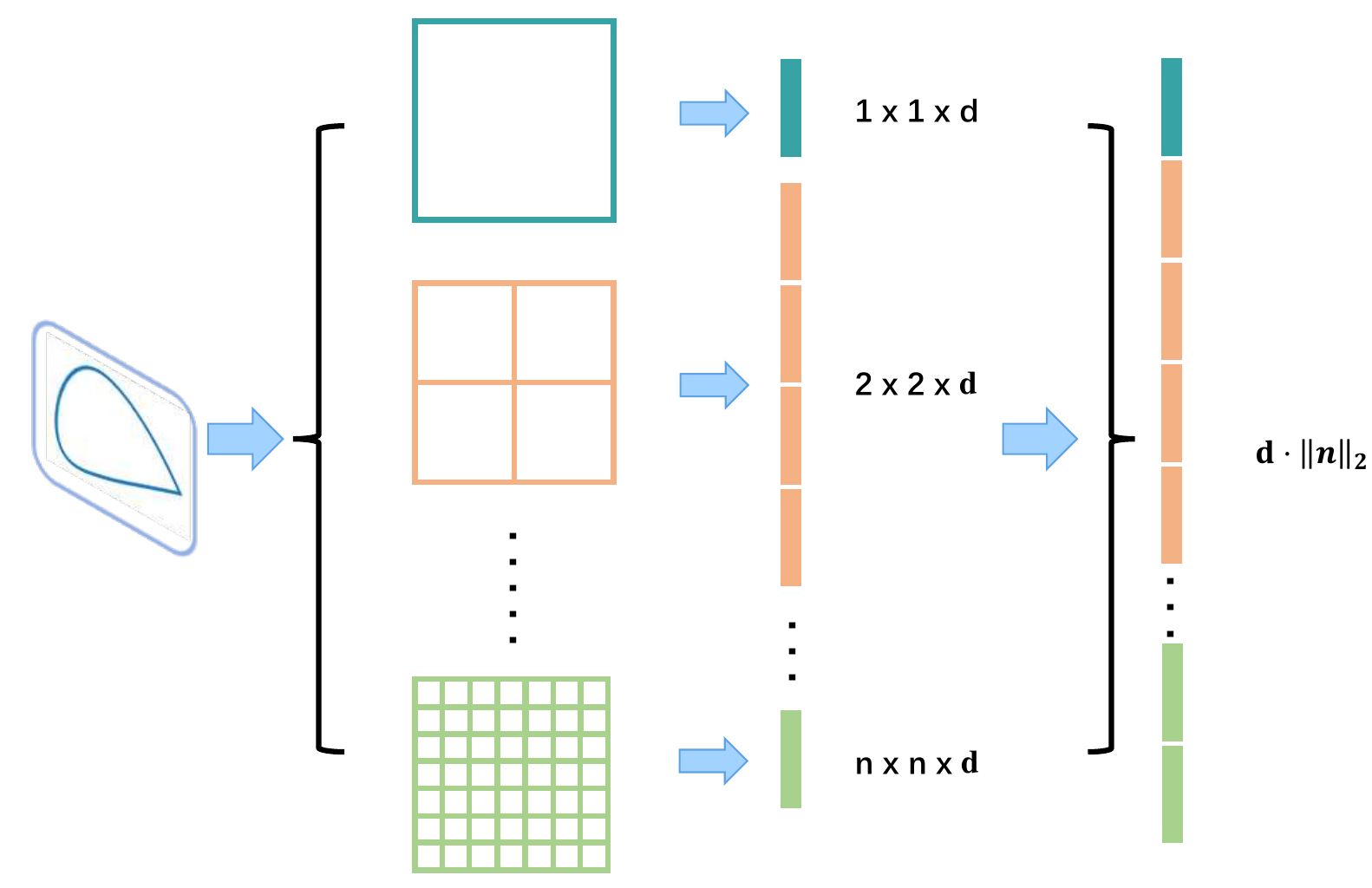}
	
	\caption{Spatial Pyramid Pooling Network.}
	
	\label{fig:sppnet}
\end{figure}

To solve this problem, we introduce spatial pyramid pooling\cite{sppnet}. We scan the feature map with a window of size $H/i \times W/i$, where $i$ is the index of pyramid layer, from 1 to $n$, and $n$ is the number of pyramid layers, it is set to 4 in our experiment. The step size of the horizontal scan is $i$ and the vertical one is $i$ as well. During the scanning process, the maximum or average value in the window is taken each step. Finally, all the results are concatenated to obtain a feature map with deterministic shape, and the shape only depends on the number of layers of the pyramid $n$. The formula is given as following:
\begin{equation}
	shape = d \cdot \sum_{i=1}^n i^2
\end{equation}
where $d$ is the number of feature map channels, $n$ is pyramid layer counts. Fig.~\ref{fig:sppnet} visualizes this process.

\subsection{Feature Extractor}

\subsubsection{Convolutional neural networks}

\begin{figure}[htbp]
	\centering
	
	\includegraphics[width=0.9\linewidth]{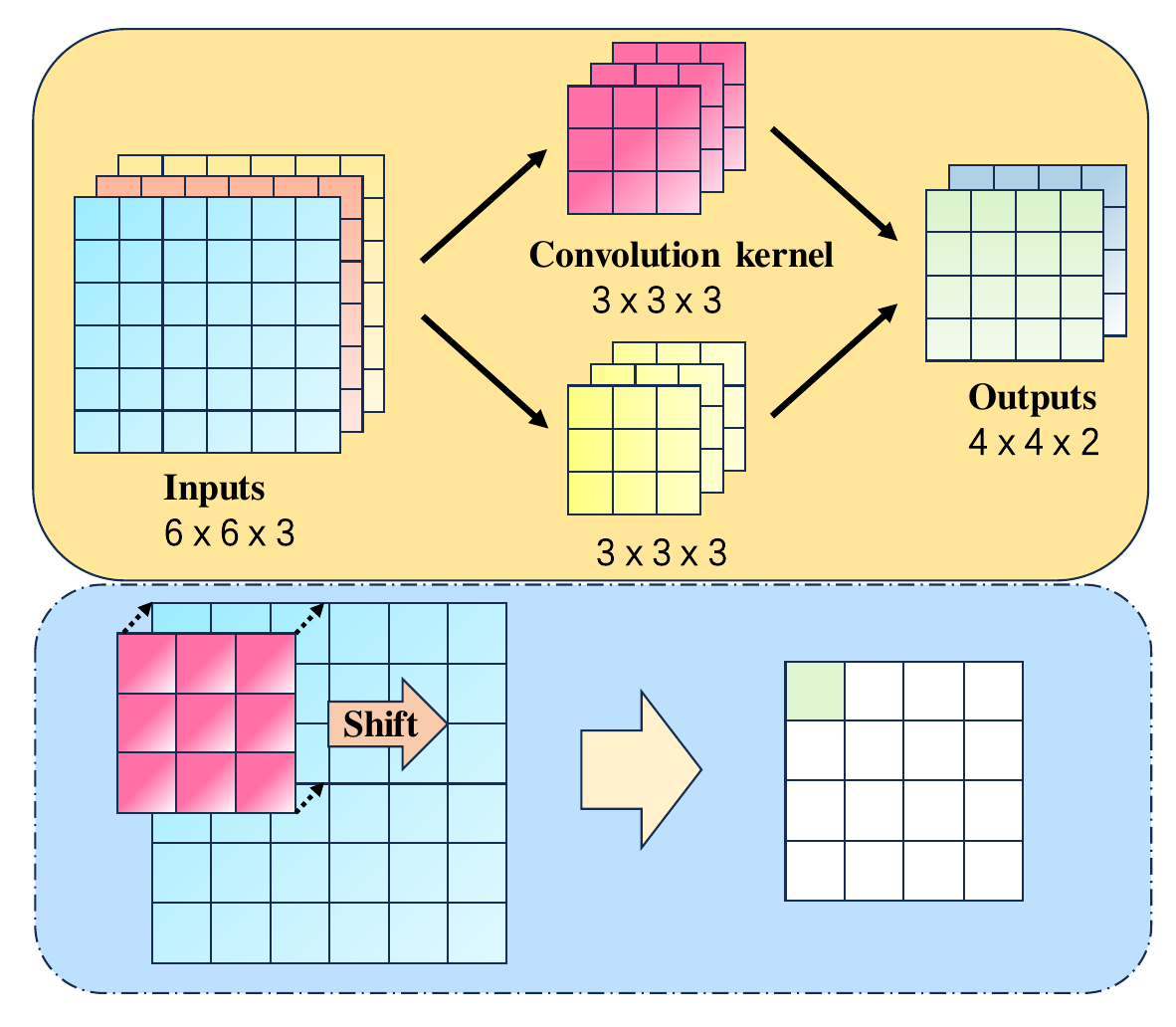}
	
	\caption{Convolutional neural networks.}
	
	\label{fig:cnn}
\end{figure}

The design of feature extractor is based on Convolutional neural networks(CNNs)\cite{cnn}. Convolutional neural networks have a great advantage in processing images. A convolutional kernel of a given size is used to scan over the image step by step. During this process, the elements in the kernel multiply with corresponding pixel elements, and the results are summed up. This action repeats until the entire image is fully scanned. Specifically, this process can be expressed as the following equation:

\begin{equation}
	O_{i, j}=\sum_{d=0}^{D-1} \sum_{m=0}^{H-1} \sum_{n=0}^{W-1} w_{d, m, n} x_{d, i+m, j+n}+w_b
\end{equation}
Where $D$ is the depth of the image, i.e., the number of channels. $H$, $W$ are the height and width of the image. $w$ is the element in the convolution kernel i.e., the training parameter, $w_b$ is the bias value and $x$ is the pixel point corresponding to the convolution kernel. Fig.~\ref{fig:cnn} visualizes this process.

\subsubsection{Max Pooling}
Typically, the image need to be downsampled to a smaller size, which improves the efficiency of the computation and allows the convolution kernel to capture richer information. In this paper, we use max pooling as a solution for downsampling. 

Similar to the convolution operation, the distinction lies in the fact that the max pooling operation selects only the pixel with the highest value among those covered by the kernel. Given a kernel of size $s \times s$, downsample the image to $H/s$, $W/s$ by setting the shift step of the kernel to $s$. Here, the size of the kernel should be the same as the shift step. It is usually set to 2.

\subsubsection{Feature extractor}
\label{sec::feature}
Our feature extractor comprises five convolutional blocks, each composed of a convolutional layer, an activation function, and a max-pooling layer. Thereby the image being downsampled to $1/32$ finally. Since the airfoil image is sparse, we set the size of the convolution kernel to $7 \times 7$ to ensure that more useful information is captured. The output channel of each convolutional layer is set to 10 in our approach.

In our study, an airfoil image with shape $H \times W \times 1$ is extracted as a feature map with shape $H/32 \times W/32 \times 10$, which is next processed by SPPNet to a uniform shape, $30 \times 10$. The feature values of the airfoil image correspond to the geometric coordinates of the airfoil, so the optimization of the feature extractor conforms to the following equation:

\begin{equation}
	\left\{
	\begin{aligned}
		& f=W_{300 \times c} \cdot x+b_{300 \times c} \\
		& L_{cnn}=\left\|g-\left(W_{c \times n} \cdot f+b_{c \times n}\right)\right\|_1
	\end{aligned}
	\right.
	\label{equ::f}
\end{equation}
where $x$ is the output of SPPNet, $f$ is the geometric coordinate feature value, $g$ is the geometric coordinate,
$L_{cnn}$ is loss function, $W$ and $b$ are the trainable parameters of the linear layer, whose subscripts are the number of neurons, $c$ is the number of geometric coordinate feature values, which is set to 2 in our method, and $n$ is the number of geometric coordinates.

\subsection{Conditional Prediction}
\label{condition}
AirfoilNet can be understood as a neural network with only the airfoil geometry coordinates as inputs, and other aerodynamic parameters such as the Reynolds number, angle of attack are used as conditions to limit the final output. This is different from the current approach. Specifically, other works\cite{fast, fast2019} input all the aerodynamic parameters in network, such as: $F (x,y,\alpha, r, p_0, p_1, \dots, p_{15})$, here $F$ is the neural network, $x$ and $y$ is the geometric coordinates of the airfoil, $\alpha$ is the angle of attack, $r$ is the Reynolds number, and $p_{i}$ is the geometric coordinate feature.

Our approach minimizes the generation of excessive trainable parameters while maintaining mathematical reliability. The Reynolds numbers are initially normalized utilizing the method described in Section~\ref{sec::encoding}. 
It is then mapped to another data distribution through an embedding layer, represented as follows:
\begin{equation}
	r^{\prime}=w_\theta r+b_{\theta},\ \ \ \  q\left(r^{\prime}\right):=\mathbb{D}_1
\end{equation}
where $w_\theta$ and $b_\theta$ denote the learnable parameters of the embedding layer, $q$ is the form of the data distribution. Alongside the angle of attack($\alpha$), $r^{\prime}$ and the geometric features($f$) jointly establish the prior condition. This process is expressed as:
\begin{equation}
	c=w_\phi \cdot cat\left(\left[\alpha, r^{\prime}, f\right]\right)+b_\phi,\ \ \ \ q\left(c\right):=\mathbb{D}_2
\end{equation}
where $c$ is the condition for prediction, $f$ is computed by Eq.~\ref{equ::f}. Finally, the geometric coordinate information is obtained by the following equation:
\begin{equation}
	g^{\prime}=w_\beta \cdot (g+c)+b_\beta,\ \ \ \ q\left(g | c\right):=\mathbb{D}_3
\end{equation}
To ensure that more information about the geometric coordinates is preserved, we employ the following approach:
\begin{equation}
	g^{\prime}=g^{\prime} + g,\ \ \ \ q\left(g^{\prime}\right):=\mathbb{D}_4
\end{equation}
This way, the prediction head solely focuses on learning the distribution of $\mathbb{D}_4$.

\subsection{Prediction Head}

\begin{figure*}[htbp]
	\centering
	
	\includegraphics[width=0.8\linewidth]{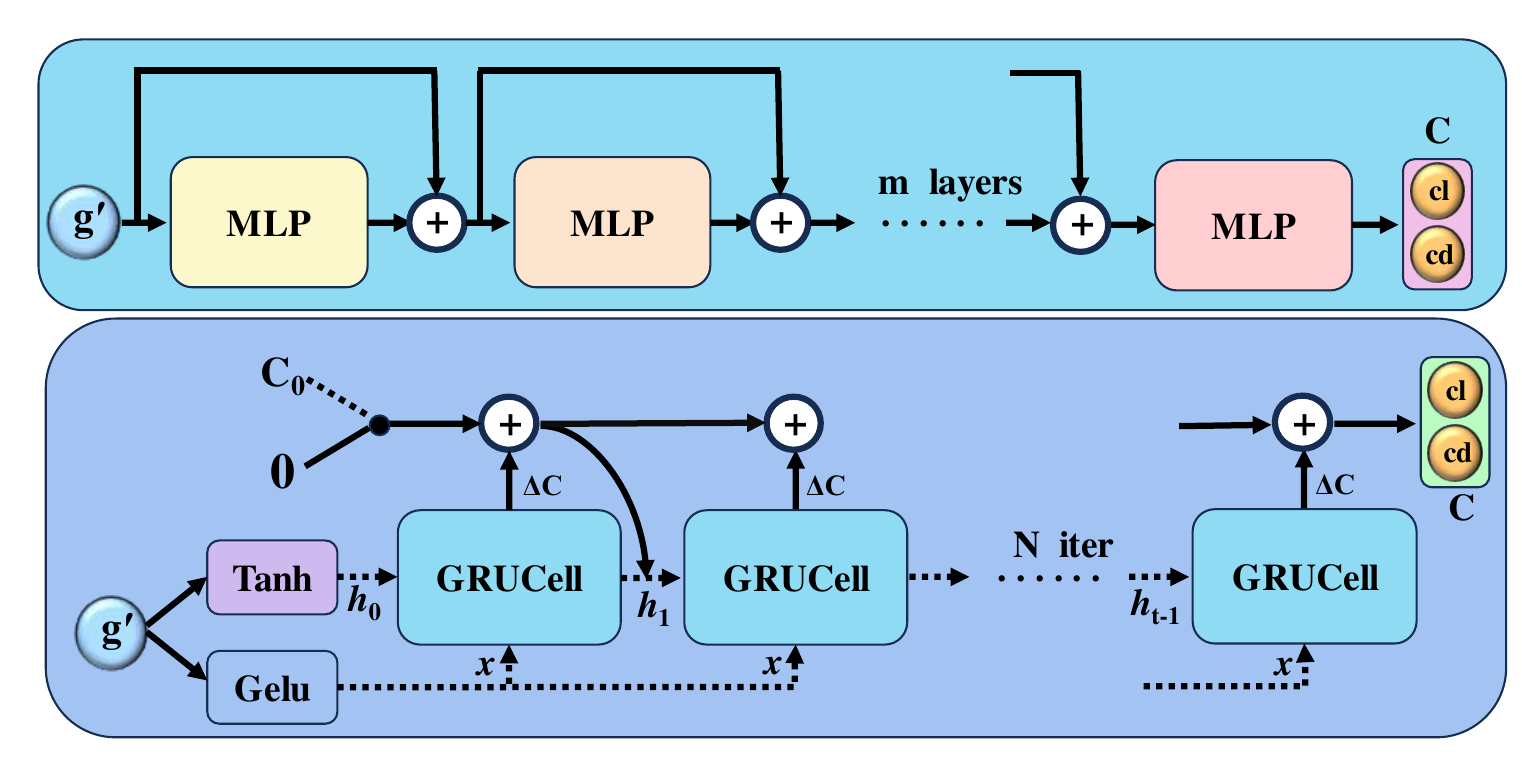}
	
	\caption{ResMLP and GRUNet. The upper half denotes ResMLP and the lower half is GRUNet.}
	
	\label{fig:prediction}
\end{figure*}

\subsubsection{ResMLP}

Existing methods currently employ very deep and wide multilayer perceptrons (MLP) to estimate aerodynamic parameters. We propose replacing this approach with a network based on Markovian properties. The details are in Fig.~\ref{fig:prediction}

In our method, the geometric coordinate information is fed into a multilayer perceptual block comprising M layers, it is set to 2 in our experiment. Each block includes five linear layers, with an activation function connected after each linear layer. To maintain gradient continuity within the -1 to 0 interval, we implement LeakyReLU as the activation function in our neural network. This consideration is based on the normalization of the Reynolds number within the range of -1 to 1. LeakyRelu is expressed as follows:
\begin{equation}
	f(x)= \begin{cases}x, & \text { if } x>0 \\ a x, & \text { otherwise }\end{cases}
\end{equation}
where a is from 0 to 1.

Following the Markov chain concept, we establish a relationship between the current state and its preceding state through summation. This approach effectively mitigates gradient vanishing issues within the ResMLP, ensuring that even with significant depth, gradients remain existed. The calculation process of ResMLP is represented as follows:
\begin{equation}
	x_{n+1}= F_n(x_n) + x_n
\end{equation}
where $F(\cdot)$ is MLP.

\subsubsection{GRUNet}
Tranditionally, the current methods output the aerodynamic parameters to be predicted directly, whereas we use a strategy, which is iterative refinement, and it makes the results more accurate. To clarify, our model consistently predicts $\Delta C$ in each iteration, culminating in the final prediction of $C$ after $N$ iterations. This idea of iterative refinement is inspired by Fan et al.'s works\cite{recloss}.

Section~\ref{condition} indicates that all inputs to the model are transformed into a distribution, denoted as $g^\prime$. It is processed with two different activation functions, yielding results that correspond to the hidden state $h_t$ and the input $x$ of the GRU, respectively. Where $h_0=tanh(g^\prime)$, $x=gelu(g^\prime)$. Here, the specific GRU algorithm is provided for the iterative update of the hidden state $h_t$:
\begin{eqnarray}
	z_t&=&\sigma(W_z\cdot[h_{t-1}, x]) \label{appa}
	\\
	r_t&=&\sigma(W_r\cdot[h_{t-1}, x]) \label{appb}
	\\
	\tilde{h_t}&=&tanh(W_h\cdot[r_t\cdot h_{t-1}, x]) \label{appc}
	\\
	h_t&=&(1-z_t) \cdot h_{t-1} + z_t \cdot \tilde{h_t} \label{appd}
\end{eqnarray}
where $W_{*}$ is learnalbe parameters, and it is shared in each iteration. This is the reason that why GRUNet only has 80k parameters. $\sigma$ is sigmoid function, $h_t$ is hidden state, and $x$ is input. The $\Delta C$ is then obtained by the following equation:
\begin{eqnarray}
	\Delta C&=&W_c \cdot h_t
\end{eqnarray}
Ultimately, C is solved for by this equation: $C_t = C_{t-1} + \Delta C$, and $C_0$ is set to 0. It is worth noting that $C_0$ can be initialized to other values, such as the output of ResMLP, which will make the final result more accurate.

\subsection{Loss Function}
The AirfoilNet is supervised based on the $l_1$ distance betweent the predicted and ground truth aerodynamic coefficients. In the ResMLP work, the loss is defined as:
\begin{eqnarray}
	L_{ResMlp}=||C_{gt} - C_{pre}||_1
\end{eqnarray}
For GRUNet, the long sequence ${C_1, C_2, \dots, C_n}$ needs to be supervised with incremental weights, computed as follows:
\begin{eqnarray}
	L_{GRUNet}=\sum_{i=1}^{N}\gamma^{N-i}||C_{gt} - C_{i}||_1
\end{eqnarray}
where N is set to 12 in our method. Since our method is end-to-end trainable, the final loss function is defined as the sum of three loss functions:
\begin{eqnarray}
	L=L_{cnn} + aL_{ResMLP} + bL_{GRUNet}
\end{eqnarray}
The training mode consists of three scenarios: first, utilizing solely ResMLP for prediction requires setting $b$ to 0; second, relying exclusively on GRUNet for prediction necessitates setting $a$ to 0; and third, using the ResMLP output as the initial value of GRUNet for prediction requires setting both $a$ and $b$ to 1.

All the learnable parameters are updated by the following equation:
\begin{eqnarray}
	w^{n+1} = w^{n} - \alpha \frac{\partial L}{\partial w} 
	\label{equ::update}
\end{eqnarray}
where $\frac{\partial L}{\partial w}$ is the gradient of the model, $\alpha$ is learning rate, and $w$ is the learnable parameters. The gradient $\frac{\partial L}{\partial w}$ is computed by chain rule, here we give a general form of this process:
\begin{eqnarray}
	\frac{\partial L}{\partial w} = \frac{\partial L}{\partial x_{out}} \times \frac{\partial x_{out}}{\partial x_{hidden}} \times\frac{\partial x_{hidden}}{\partial w} 
\end{eqnarray}
For example, there is an target function $y=ax+b$, and the deep learning model $y^*=w_{\theta}x+w_b$ is going to fit this curve with mean square error: $(y-y^*)^2$. According to the chain rule, the gradient of $y^*$ can be obtained by the following:
\begin{eqnarray}
	\begin{cases}
		\frac{\partial L}{\partial w_\theta }= \frac{\partial L}{\partial y^*}\times \frac{\partial y^*}{\partial w_\theta }=2(y-y^*)x
		\\
		\frac{\partial L}{\partial w_b}= \frac{\partial L}{\partial y^*}\times \frac{\partial y^*}{\partial w_b }=2(y-y^*)
		
	\end{cases} 
\end{eqnarray}
Eq~\ref{equ::update} then allows $w_\theta$ to approximate $a$, while complex deep learning models can be viewed as pushing the general case to higher dimensions.

\section{EXPERIMENT}
\subsection{Data Preparation}
AirfoilNet is trained on a public datasets\cite{data}. It contains the main aerodynamic characteristics of airfoil shapes for aircraft wings developed by the National Advisory Committee for Aeronautics (NACA). We have opted for 4 and 5-digit airfoil for experiments. There are 20 types of the 4-digit airfoil and 5 types for the 5-digit airfoil.
In our experiments, one of the 4-digit airfoils is used as a validation set to evaluate the model. Each airfoil data includes 5 Reynolds numbers(50k, 100k, 200k, 500k, 1000k), each of which corresponds to the lift and drag coefficients at 90 angles of attack(from -15.5 to 19).
Thus a total of 10264 data($<10800$, some types of AOA are less than 90) are used for training and 637 for validation.

\subsection{Training Schedule}
AirfoilNet is implemented in PyTorch\cite{pytorch} and trained using tow 2080Ti GPUs. We trained AirfoilNet with three different prediction heads. Specifically, during the training phase of ResMLP, we opted for a batch size of 256 and set the learning rate to 2.5e-4. When transformed to GRUNet, the batch size is set to 512 and learning rate is increased to 4e-4. If the two are trained in combination, the batch size is set to 256, learning rate is 2.5e-4 as well.All training processes contain 1,000 epochs, and the model weights are initialized randomly and identically for each training process.

\subsection{Results and Discussions}
There are two metrics be used to evaluate our model, which are the mean squared error (MSE) and the relative error (RE) between the predicted and true values. They are defined as follows: 

\begin{eqnarray}
	MSE = \frac{1}{2N}\sum_{i=1}^{N} [(Cl^{i}_{pre} - Cl^{i}_{t})^2+(Cd^{i}_{pre} - Cd^{i}_{t})^2]
\end{eqnarray}

\begin{eqnarray}
	\begin{cases}RE_{Cl} =& \frac{1}{N} \sum_{i}^{N}\left |\frac{Cl^i_{pre} - Cl^i_{t}}{Cl^i_{t}}\right |
		\\
		RE_{Cd} =& \frac{1}{N} \sum_{i}^{N}\left |\frac{Cd^i_{pre} - Cd^i_{t}}{Cd^i_{t}}\right |
	\end{cases}
\end{eqnarray}

where N is batch size, $C_{pre}$ is the predicted value, $C_{t}$ is ground truth.

\subsubsection{Error analysis}
The Table~\ref{tab::results} shows that the effect of different prediction head on the results, and these data are obtained from the evaluation of AirfoilNet on NACA 2424 airfoil. Upon receiving the initial values from GRUNet, AirfoilNet demonstrates a marked 55.7\% improvement in its performance. Due to its non-iterative nature, ResMLP incurs minimal computational overhead. Compared to GRUNet, the time overhead has decreased by 40\% with ResMLP. The inference time in Table~\ref{tab::results} represents the time that the model infer on the GPUs. Meanwile, we compare inference times on CPUs as well. The results are presented in Table~\ref{tab::time}.


\begin{table}[h]
	\centering
	
	\begin{tabular}{cccccccclc}
		\hline
		\multirow{2}{*}{Head} &  & \multirow{2}{*}{MSE} &  & \multicolumn{2}{c}{RE} & \multicolumn{1}{l}{} & \multirow{2}{*}{Time(s)} & \multicolumn{1}{c}{} & \multirow{2}{*}{Parameters(K)} \\ \cline{5-6}
		&  &                      &  & Cl         & Cd        & \multicolumn{1}{l}{} &                                    &                      &                                \\ \hline
		GRUNet                    &  & 8.24e-3              &  & 0.3655     & 0.1485    &                      & 0.010                              &                      & 80                             \\
		ResMLP                 &  & 8.30e-3              &  & 0.3377     & 0.1893    &                      & 0.006                              &                      & 120                            \\
		GRUNet$^*$                    &  & 3.65e-3              &  & 0.3814     & 0.0999    &                      & 0.010                              &                      & 160                            \\ \hline
	\end{tabular}
	
	\captionsetup{singlelinecheck=false, justification=raggedright}
	\caption{The resultes of MSE and RE in different predicted methods. GRUNet$^*$ indicates the result when it is initialized to the output of ResMLP.}
	\label{tab::results}
\end{table}

\begin{figure*}[h]
	\centering
	\begin{minipage}[b]{1\textwidth}
		\centering
		\includegraphics[width=6cm]{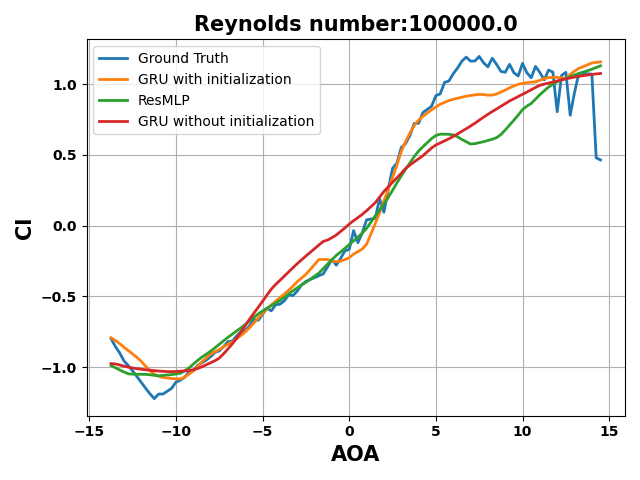}
		\includegraphics[width=6cm]{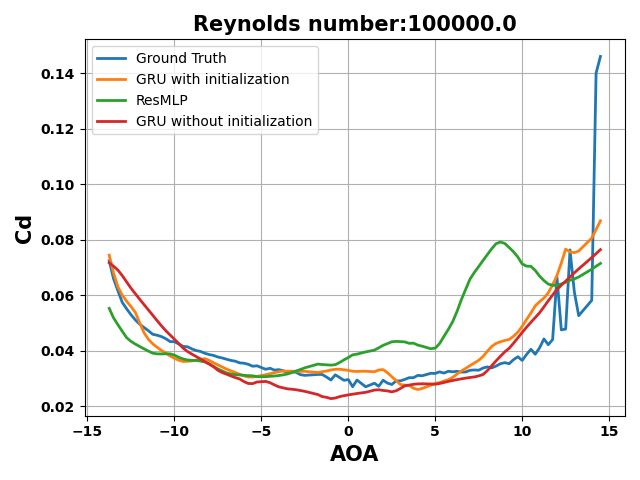}

	\end{minipage}
	\hfill
	\begin{minipage}[b]{1\textwidth}
		\centering
		\includegraphics[width=6cm]{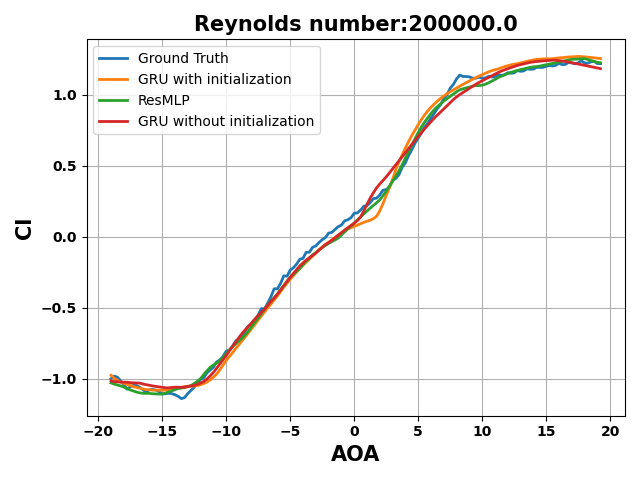}
		\includegraphics[width=6cm]{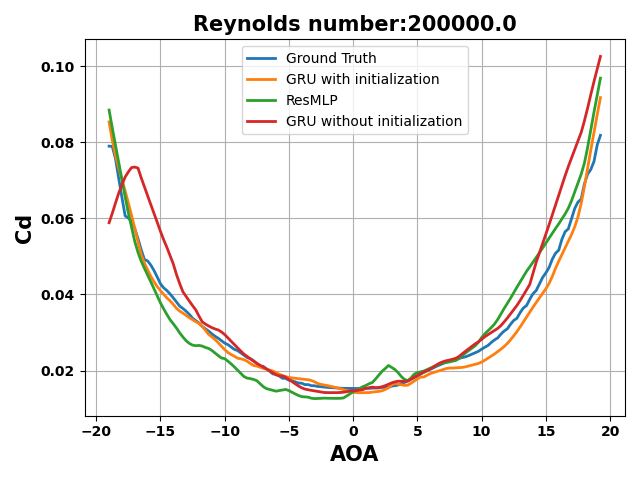}

	\end{minipage}
	\hfill
	\begin{minipage}[b]{1\textwidth}
		\centering
		\includegraphics[width=6cm]{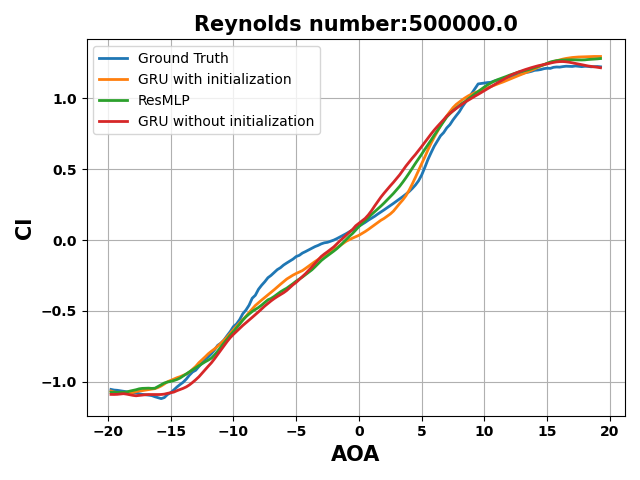}
		\includegraphics[width=6cm]{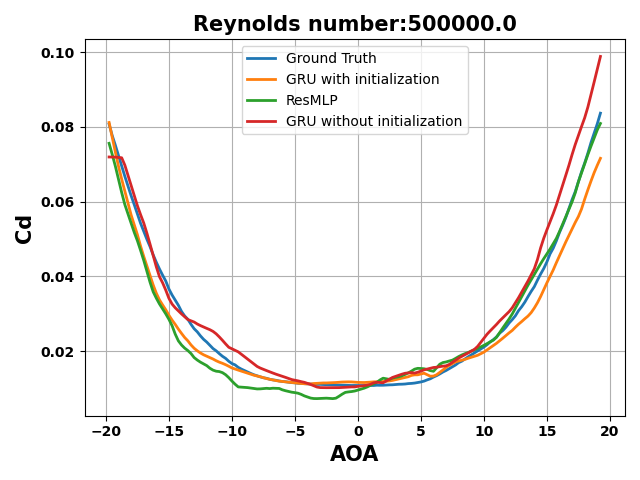}
		\label{fig:image2}
	\end{minipage}
	\\[10pt]
	\begin{minipage}[b]{1\textwidth}
		\centering
		\includegraphics[width=6cm]{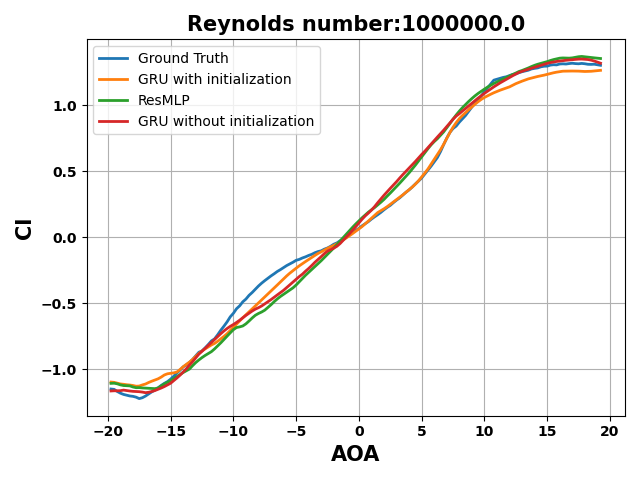}
		\includegraphics[width=6cm]{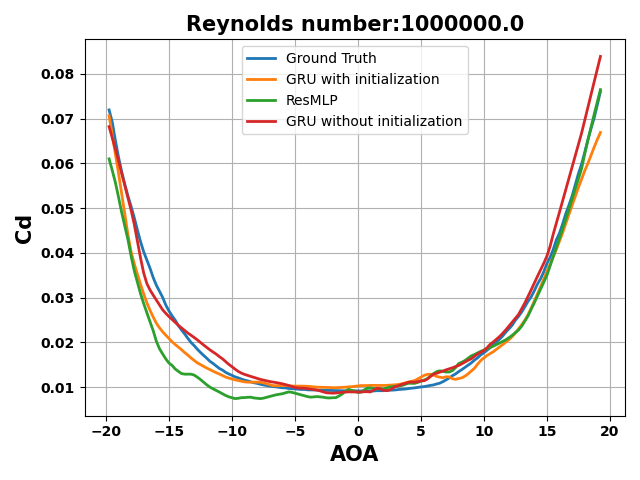}

		\label{fig:image3}
	\end{minipage}
	\hfill
\caption{Error Curve. Ground Truth is the reference value.}
\label{fig::curve}
\end{figure*}

Figure~\ref{fig::curve} presents the overall error visualization plot, where the reference, denoting the ground truth, is depicted in blue. It illustrates the predictive performance of various prediction heads in AirfoilNet.

\begin{table}[]
	\centering
	\resizebox{0.4\textwidth}{!}{%
	\begin{tabular}{cclclcl}
		\cline{1-6}
		\multicolumn{6}{c}{Inference Time(s)}                  &  \\ \cline{1-6}
		\multicolumn{1}{l}{} & ResMLP &  & GRU   &  & GRU$^{*}$   &  \\
		CPU                  & 0.008  &  & 0.013 &  & 0.013 &  \\
		GPU                  & 0.006  &  & 0.010 &  & 0.010 &  \\ \cline{1-6}
	\end{tabular}}
	\captionsetup{singlelinecheck=false, justification=raggedright}
	\caption{Inference time of different prediction heads in AirfoilNet on various devices.}
	\label{tab::time}
\end{table}

\subsubsection{Comparison experiment}
Firstly, we compared the impact of utilizing or omitting the Reynolds number encoding scheme on the prediction results. This experiment was conducted on the ResMLP prediction head, and in fact our method is more effective when the Reynolds number is denser and the maximum-minimum gap is more drastic.

\begin{table}[h]
	\centering
	\begin{tabular}{clclllll}
		\hline
		\multirow{2}{*}{Head} &  & \multirow{2}{*}{Reynolds number encoding} &  & \multicolumn{1}{c}{\multirow{2}{*}{MSE}} &  & \multicolumn{2}{c}{RE}                          \\ \cline{7-8} 
		&  &                                           &  & \multicolumn{1}{c}{}                     &  & \multicolumn{1}{c}{Cl} & \multicolumn{1}{c}{Cd} \\ \hline
		\multirow{2}{*}{ResMLP}          &  & $\checkmark$                                         &  & 8.30e-3                                  &  & 0.3377                 & 0.1893                 \\
		&  & ×                                         &  & 1.18e-2                                  &  & 0.4159                 & 0.2321                 \\ \hline
	\end{tabular}
\captionsetup{singlelinecheck=false, justification=raggedright}
\caption{Impact of Reynolds number encoding on the model.}
\end{table}

We proceeded by configuring various ResMLP layer counts to evaluate their respective impacts on outcomes. Due to parameter sharing across every ResMLP layer, augmenting the layer count does not result in an increase in parameter quantity. Table~\ref{tab::layers} summarizes all results. It illustrates that an increase in the number of layers leads to improved performance of the model.

\begin{table}[h]
	\begin{tabular}{clclll}
		\hline
		\multirow{2}{*}{Head}  &  & \multicolumn{1}{l}{\multirow{2}{*}{The number of layer}} &  & \multicolumn{2}{c}{RE}                          \\ \cline{5-6} 
		&  & \multicolumn{1}{l}{}                        &  & \multicolumn{1}{c}{Cl} & \multicolumn{1}{c}{Cd} \\ \hline
		\multirow{3}{*}{ResMLP} &  & 2                                           &  & 0.3377                 & 0.1893                 \\
		&  & 4                                           &  & 0.2945                 & 0.0931                 \\
		&  & 6                                           &  & 0.2589                 & 0.1285                 \\ \hline
	\end{tabular}
\caption{Effect of multiple layers of ResMLP on results.}
\label{tab::layers}
\end{table}

We conducted a comparative analysis of the outcomes derived from employing various pooling operations within the spatial pooling pyramid.For a grayscale image representing a wing, where pixel values are binary (0 and 1), the use of max pooling presents limited generality due to the inherent restriction of the maximum value within a window to 1. Intuitively, to enhance the model's generalization capability, employing mean pooling seems more appropriate. This is confirmed by the results in Table~\ref{tab::pool}.

\begin{table}[]
	\begin{tabular}{llllllllcl}
		\hline
		\multicolumn{1}{c}{\multirow{2}{*}{Head}} &  & \multirow{2}{*}{Layers}                &  & \multirow{2}{*}{Pooling in SPPNet} &  & \multicolumn{1}{c}{\multirow{2}{*}{MSE}} &  & \multicolumn{2}{c}{RE}          \\ \cline{9-10} 
		\multicolumn{1}{c}{}                      &  &                                        &  &                                    &  & \multicolumn{1}{c}{}                     &  & Cl     & \multicolumn{1}{c}{Cd} \\ \hline
		\multirow{2}{*}{ResMLP}                   &  & \multicolumn{1}{c}{\multirow{2}{*}{2}} &  & \multicolumn{1}{c}{Max}            &  & 8.30e-3                                  &  & 0.3377 & 0.1893                 \\
		&  & \multicolumn{1}{c}{}                   &  & \multicolumn{1}{c}{Mean}           &  & 5.62e-3                                  &  & 0.3294 & 0.0967                 \\ \hline
	\end{tabular}
\captionsetup{singlelinecheck=false, justification=raggedright}
\caption{Impact of pooling operations on results in SPPNet.}
\label{tab::pool}
\end{table}

Finally, we validated the advantages of the iterative refinement scheme. It facilitates AirfoilNet to converge consistently even after numerous iterations during the inference phase. \ref{fig::iterations} provides a visualization. This also signifies that the outcomes $\Delta C$ from each update of GRUNet consistently converge to a fixed point.

\begin{minipage}{\linewidth}
		\centering
		\includegraphics[width=\linewidth]{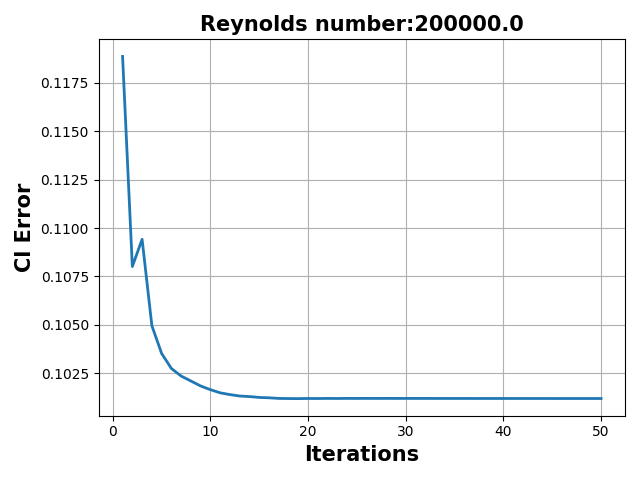}
	\includegraphics[width=\linewidth]{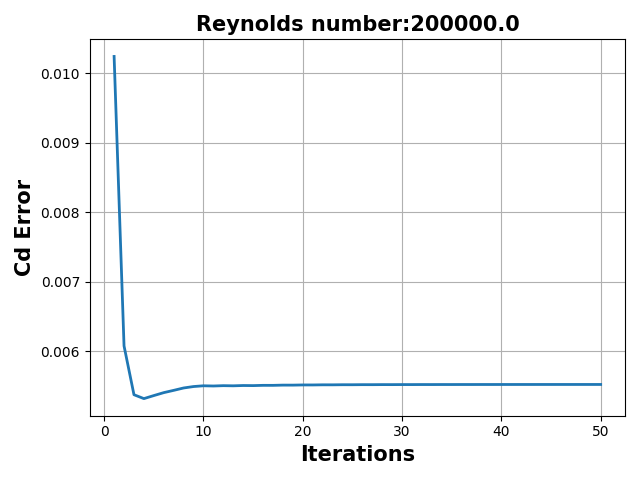}
	\captionof{figure}{Lift and drag coefficient errors as a function of the number of iterations at inference time.}
	\label{fig::iterations}
\end{minipage}

\section{Conclusions}
In this paper, we propose a method called AirfoilNet to predict the lift and drag coefficients using deep learning techniques. The method addresses the shortcomings present in existing approaches, such as the inability to perform end-to-end training and prediction, as well as larger parameter sizes. AirfoilNet incorporates two different prediction heads, ResMLP, GRUNet, which are designed for our proposed new prediction strategy. The strategy allow models to achieve more accurate resultes.

Our approach can be seen as a tool accessible to non-professionals, designed for easy utilization. Upon inputting the airfoil image and its corresponding geometric coordinates, subsequent specification of the angle of attack and Reynolds number facilitates the derivation of the respective lift and drag coefficients. Our approach exhibits remarkable scalability, and we envision future researchers employing AirfoilNet for predicting other aerodynamic parameters. We aim to explore a methodology in the future to address the scarcity of training data. This plays a pivotal role in data-driven methodologies.

\begin{acknowledgments}
	We wish to acknowledge the support of the author community in using
	REV\TeX{}, offering suggestions and encouragement, testing new versions,
	\dots.
\end{acknowledgments}

\nocite{*}
\bibliography{aipsamp}

\end{document}